\documentclass[11pt]{article}
\pdfoutput=1
\usepackage{jheppub}
\usepackage{amssymb,amsfonts}
\usepackage{mathrsfs}
\let\savenumberline\numberline
\def\numberline#1{\savenumberline{#1.}}
\makeatletter
\renewcommand{\@seccntformat}[1]{\csname the#1\endcsname.\,\,}
\makeatother

\newcommand{\R}{{\bf R}}

\newcommand{\CH}{{\cal H}}

\newcommand{\CL}{{\cal L}}
\newcommand{\CM}{{\cal M}}

\newcommand{\p}{\partial}
\renewcommand{\tilde}[1]{\widetilde{#1}}
\renewcommand{\hat}[1]{\widehat{#1}}
\newcommand{\be}{\begin{equation}}
\newcommand{\ee}{\end{equation}}
\newcommand{\bea}{\begin{eqnarray}}
\newcommand{\eea}{\end{eqnarray}}

\makeatletter
\def\@fpheader{\relax}
\makeatother
\usepackage{graphicx}
\usepackage{latexsym}
\newcommand{\D}{{\textrm{D}}}
\title{\ \vspace{1.5in} \\ \hbox{Covariant Hamilton-Jacobi Equation for Pure Gravity}}
\author{Petr Ho\v{r}ava}
\affiliation{\medskip
Berkeley Center for Theoretical Physics and Department of Physics\\
University of California, Berkeley, CA, 94720-7300, USA\medskip\\
Theoretical Physics Group, Lawrence Berkeley National Laboratory\\
Berkeley, CA 94720-8162, USA}
\abstract{The main purpose of this article is to provide access to a previously unpublished and nearly lost paper: P. Ho\v{r}ava, \textit{Covariant Hamilton-Jacobi Equation for Pure Gravity}, which appeared originally in July 1990 as a Prague Preprint PRA-HEP-90/4, at the Institute of Physics, Czechoslovak Academy of Sciences, but appears otherwise unavailable online.  The author has recently acquired an original copy of this preprint; the present article contains a verbatim transcript of the original 1990 paper, framed by a small number of comments.  The contents of the 1990 paper was based on the results contained in the author's BSc.\ Thesis, written in Czech, and presented at Charles University, Prague, in 1986.}
\begin{document}
\notoc
\maketitle\vfill\break
%%%%%%%%%%%%%%%%%%%%%%%%%%%%%%%%%%%%%%%%%%%%%%%%%%%%%%%%%%%%%%%%%%%%%%%%%%%%%%%
\begin{center}
  \textbf{Preamble}
\end{center}
\smallskip
This article contains the transcript of an original copy of a previously unpublished paper: P. Ho\v{r}ava, \textit{Covariant Hamilton-Jacobi Equation for Pure Gravity}, which first appeared in July 1990 as a Prague Preprint PRA-HEP-90/4, at the Institute of Physics, Czechoslovak Academy of Sciences, where I was a graduate student at the time.  This preprint is not only unpublished, but appears unavailable anywhere online.  During the past three decades, I have been receiving a steady trickle of requests for a copy of the original 1990 preprint; only very recently I have acquired an original copy, which allows me to make it finally publicly available in the form of this arXiv posting.
\smallskip

This Preamble is followed by the faithful transcript of the original 1990 paper, verbatim with only one exception:  All the cited references have been modernized to the currently accepted format, and all the references to articles in their preprint form have been updated to include the full information about their final published version.  I have resisted the temptation to add any additional references to works written since this preprint appeared in 1990, and/or to provide any additional comments on the more recent scientific developments in this field.  The transcribed version of the 1990 paper is then followed by the scanned 10 pages of the original copy of the 1990 preprint.  
\smallskip

The 1990 paper is substantially based on the geometric language and structures of the covariant calculus of variations on fibered manifolds and their jet prolongations.  For the readers who wish to consult some more general background information on this modern covariant geometric approach to the calculus of variations and the geometry of jet prolongations of fibered manifolds, we recomment the overview book by D.~Krupka, \textit{Introduction to Global Variational Geometry} (Atlantis Press, Paris, 2015), and the references therein.  (Additional details related to the 1990 paper are also provided in Ref.~[9] of the transcribed version.)
\smallskip

The results of this 1990 paper (and the published sequel, cited in the transcribed version as Ref.~[9]) were entirely based on my BSc.\ Thesis, submitted and defended in 1986 at Charles University, Prague, with Prof.\ Demeter Krupka (then of Masaryk University, Brno) as my BSc.\ Thesis advisor.  I am grateful to Prof.\ Demeter Krupka and to Prof.~Ji\v{r}\'\i~Niederle (my PhD.\ advisor at the Institute of Physics, Prague), for their invaluable guidance and discussions during the time span of the years leading to my 1990 paper as transcribed here.  

\vfill\break
%%%%%%%%%%%%%%%%%%%%%%%%%%%%%%%%%%%%%%%%%%%%%%%%%%%%%%%%%%%%%%%%%%%%%%%%%%%%%%%
\noindent
\textit{Institute of Physics, Czechosl.\ Acad.\ of Sci.}\hfill\textbf{PRA-HEP-90/4}\break
\textit{and}\hfill\textbf{July 1990}\break
\textit{Nuclear Centre, Charles University}\hfill\break
\textit{Prague}

\vglue1in

\begin{center}{\huge\bf{Covariant Hamilton-Jacobi Equation\\ \smallskip for Pure Gravity}${}^\ast$}

\bigskip\bigskip

{\Large\sc{Petr Ho\v{r}ava}${}^\dagger$}
\bigskip

\textit{Institute of Physics \v{C}SAV}
\smallskip

\textit{CS-18040 Prague 8}
\smallskip

\textit{Czechoslovakia}
\bigskip\bigskip

\textbf{Abstract}
\end{center}
\medskip

\noindent We present an alternative framework for treating Einstein gravity in any dimension greater than two, and at any signature.  It is based on a covariant Hamilton-Jacobi-De~Donder equation, which is proved to be equivalent to the Lagrange theory, on space-times of arbitrary topology.  It in particular means that Einstein gravity can be thought of as a (covariantly) regular system.  Finally, the Hamilton-Jacobi theory is studied, and it is shown that any solution of Einstein equations can be obtained from the action form equal identically to zero.
\medskip

\noindent
PACS numbers: 04.20.Fy, 04.20.Cv, 04.60.+n.\vfill

\noindent\small{${}^\ast$Revised version: November 1990}\hfill\break
\noindent\small{${}^\dagger$e-mail address: {\tt{horava@cspgas11.bitnet}}\hfill
  \pagebreak
%%%%%%%%%%%%%%%%%%%%%%%%%%%%%%%%%%%%%%%%%%%%%%%%%%%%%%%%%%%%%%%%%%%%%%%%%%%%%%%
  \baselineskip15pt
\renewcommand\theequation{\arabic{equation}}

  1. The problem of reconciling classical gravity with quantum theory still resists the effort done in this field.  It is not only the fact that Einstein gravity is not renormalizable in more than three dimensions that prevents a substantial progress in quantum gravity.  Some other conceptual problems remain to be clarified before we get an ultimate quantum version of the theory of gravitation.
  \smallskip
  
  It might well be that gravity can only be consistently quantized within a unified theory, perhaps some deeper version of string theory.  Nevertheless, some of the crucial problems are already present in the classical formulation of pure Einstein gravity.  One of the most important and challenging is the problem of time in quantum gravity, stemming essentially from the fact that the r\^{o}le of time in general relativity is quite different from the r\^{o}le it plays in quantum physics.  With the desired unifying theory being still far from us, we shall concentrate in this Letter on pure Einstein gravity with its scalar curvature Lagrangian, and shall try to throw some new light on its structure.
  \smallskip

  To have a well-understood Hamilton and Hamilton-Jacobi formulation of the theory of gravitation is of utmost importance for its quantization.  Bearing in mind the problems with time in quantum gravity, the standard way to construct a Hamilton (or Hamilton-Jacobi) theory for gravity is to slice the space-time into a $3+1$ dimensional foliation, and consider evolutionary Hamilton equations, as \textit{e.g.} in the ADM formalism.  We shall present here another version of Hamilton and Hamilton-Jacobi descriptions of gravity that, in our opinion, has some advantages (and presumably some disadvantages) when compared with the standard versions of Hamiltonian gravity.  In particular, the formulation presented here does not presume any preferred r\^{o}le of time.
  \smallskip

  Our framework stems from the covariant variational calculus, with its own geometrical generalization of the Hamilton and Hamilton-Jacobi theories of mechanics to field theory.  Nevertheless, because the results are essentially independent of their variational origin, we shall try to explain them without excessive geometry.
  \medskip

  2.  Let us first fix some notation.  Gravitational field on a given space-time manifold $\CM$ with coordinates $(x^a)$ will be described (in order to simplify some formulas) by the contravariant tensor density with components $h^{ab}=eg^{ab}$, where $g_{ab}$ is the conventional metric tensor, and $e$ is a shorthand for $\sqrt{|\det g_{ab}|}$. All of the results of this letter are valid in any space-time dimension $D>2$, and for any metric signature.  Thus, the indices $a,b,\ldots$ go from 1 to $\D$.
  \smallskip

  Whereas the Hamilton-Jacobi equation of mechanics is an equation for one unknown function $S$ of coordinates $q^\sigma$ and time $t$, the proposed Hamilton-Jacobi equation for gravity is an equation for $\D$ unknown functions $S^a$ of metric components $h^{ab}$ and space-time coordinates $x^a$:
\be
\frac{\p S^a}{\p x^a}+h^{ac}\left\{\frac{\p S^d}{\p h^{ab}}\frac{\p S^b}{\p h^{cd}}+\frac{1}{(1-\D)}\frac{\p S^b}{\p h^{ab}}\frac{\p S^d}{\p h^{cd}}\right\}=0.
\label{eeone}
\ee
Were we considering an analogous covariant HJ equation for simpler field systems (as \textit{e.g.} a scalar field theory), the $S^a$'s would represent components of a differential $(D-1)$-form $S$.  For gravity instead, the functions $S^a$ of Eq.~(\ref{eeone}) comprise a differential form $Z$ in a more complicated manner:
\be
Z=\sum_aZ^a\omega_a,\qquad Z^a=S^a-\frac{\p S^a}{\p h^{bc}}h^{bc},
\label{eetwo}
\ee
where $\omega_a=(-1)^{a+1}dx^1\wedge\ldots\wedge \hat{dx^a}\wedge\ldots\wedge dx^D$. From Eq.~(\ref{eetwo}), the following transformation properties result for $S$ with respect to a coordinate change $\tilde x^a=\tilde x^a(x^b)$:
\be
\tilde S^a=\det\left(\frac{\p x}{\p\tilde x}\right)\left\{\frac{\p\tilde x^a}{\p x^e}S^e+\frac{\p^2\tilde x^a}{\p x^b\p x^c}h^{bc}-\frac{\p\tilde x^a}{\p x^b}\frac{\p x^e}{\p\tilde x^d}\frac{\p^2\tilde x^d}{\p x^e\p x^c}h^{bc}\right\}.
\nonumber
\ee
The fact that $S$ is a geometrical object with such perculiar transformation properties is closely related to the principle of equivalence.
\smallskip

The basic claim of this letter is that the whole dynamics of the classical gravitational field is hidden in Eq.~(\ref{eeone}).  More explicitly, to find a solution of the Einstein equations, we must proceed in two steps.  First, we have to find a solution of the Hamilton-Jacobi equation (\ref{eeone}), \textit{i.e.}, to find some functions $S_0^a$ of $x^b$ and $h^{cd}$ that annihilate the left-hand side of Eq.~(\ref{eeone}).  With such functions $S_0^a$ at hand, one constructs the following system of first-order equations:
\be
\frac{\p\bar h^{bc}}{\p x^a}=\bar h^{d(b}\frac{\p S_0^{c)}}{\p h^{ad}}(x,\bar h)+\frac{1}{(1-\D)}\delta_a^{(b}\bar h^{c)d}\frac{\p S_0^e}{\p h^{de}}(x,\bar h)
\label{eethree}
\ee
for unknown functions $\bar h^{bc}$ of $x^a$.  (Our conventions throughout the Letter are such that, \textit{e.g.}, $\delta_a^{(b}\bar h^{c)d}=\delta_a^{b}\bar h^{cd}+\delta_a^{c}\bar h^{bd}$.) Every solution of Eq.~(\ref{eethree}) solves the Einstein equations.  The opposite statement is true as well, \textit{i.e.}, every solution of the Einstein equations can be obtained this way from an appropriate solution of the Hamilton-Jacobi equation (see below).
\smallskip

Thus, within the Hamilton-Jacobi theory, the process of solving the Einstein equations has been divided into two consecutive steps, each of them looking -- at least naively -- more tractable than the Lagrange set of the Einstein equations.  In particular, the problem of integrating the set of first-order differential equations (\ref{eethree}) is clearly equivalent to searching for integrals of a $\D$-dimensional vector distribution (in the sense of Frobenius) generated by vector fields $\xi_{(a)}$,
\be
\xi_{(a)} =\frac{\p}{\p x^a}+\left\{ h^{d(b}\frac{\p S_0^{c)}}{\p h^{ad}}+\frac{1}{(1-\D)}\delta_a^{(b} h^{c)d}\frac{\p S_0^e}{\p h^{de}}\right\}\frac{\p}{\p h^{bc}},
\label{eefour}
\ee
where $a=1,\ldots, \D$.  Integration of vector distributions is a well-known problem in differential geometry.  Unfortunately, the distribution (\ref{eefour}) is usually (\textit{i.e.}, for generic $S_0^a$) not completely integrable, and we have to look for its individual integral hypersurfaces.  After all that, we can obtain global solutions by sewing the local patches together.
\medskip

3.  The key to the Hamilton-Jacobi equation and its equivalence to the Lagrange framework, is the geometrical version of variational calculus as presented \textit{e.g.} in Refs.~\cite{dedonder,gs,krupka,franca}.  We shall only sketch here some crucial facts concerning the geometry, referring the interested reader to the literature for more details on the general variational calculus.
\smallskip

The central object of the theory is the so-called Lepagean equivalent \cite{krupka} of given Lagrangian $\CL$, which is a differential $\D$-form, equivalent to the Lagrangian when used to evaluate the action on field configurations, but differing from it by terms that allow one to geometrize the first variational formula in terms of differential forms.  To be more specific, the Lepagean equivalent of classical mechanics is identical to the well-known Poincar\'e-Cartan form \cite{cartan}.
\smallskip

The Lepagean equivalent underlies the geometric version of Hamilton theory.  We can associate naturally with a given Lepagean equivalent a new variational problem with its own Lagrangian.  The Lagrange equations of this new Lagrangian are first-order differential equations, which treat fields as well as their derivatives up to the order the Lepagean equivalent depends on, as independent variables.  This is in close analogy with taking coordinates and momenta as independent variables in Hamiltonian mechanics, thus motivating why the equations are referred to as the Hamilton-De~Donder equations.  Under some assumptions of regularity of $\CL$, they are equivalent to the set of the Lagrange equations of $\CL$.  In essence, the Hamilton-De~Donder equations represents the first order formalism of field theory.
\smallskip

For generic second-order Lagrangians, the Lepagean equivalent depends on the derivatives of fields up to the third order.  The first problem we are faced with in the case of pure gravity is the fact that the Lagrangian is badly singular in the above-mentioned sense as a second-order lagrangian.  This problem is solved by the simple but crucial observation \cite{ks} that, as $\CL$ is of first order up to a total derivative, the Lepagean equivalent essentially depends upon the first derivatives of $h^{ab}$.  Explicitly, the Lepagean equivalent $\rho$ of the gravitational Lagrangian $eR\,dx^1\wedge\ldots \wedge dx^\D\equiv eR\,\omega_0$, can be written in coordinates as
\be
\rho=\theta+d\sigma,
\label{eefive}
\ee
where
\bea
\theta&=&h^{ac}\Gamma_{a[b}^d\Gamma_{c]d}^b\omega_0+Q_{bc}^a\left(dh^{bc}-h^{bc}{}_{,d}dx^d\right)\wedge\omega_a,\nonumber\\
\sigma&=&-Q_{bc}^ah^{bc}\omega_a,\nonumber\\
Q_{bc}^a&=&\frac{1}{2}\delta_{(b}^a\Gamma_{c)d}^d-\Gamma_{bc}^a.
\label{eesix}
\eea
We are now free to apply the Hamilton-De~Donder strategy to the system as if it were of first order.  In this sense, pure gravity is a regular system, and the Hamilton-De~Donder theory represents the first order formulation, written in variables $(h^{ab},Q_{bc}^a)$.
\smallskip

As for the Hamilton-Jacobi equation, its geometrical version reads
\be
w^\ast d\rho=0,
\label{eeseven}
\ee
with unknown mapping $w$, which in coordinates gives $Q_{bc}^a$ as functions of $(x^d,h^{de})$. Clearly, this equation is locally equivalent to
\be
w^\ast \rho=dZ,
\label{eeeight}
\ee
where $Z=Z^a(x^b,h^{cd})\omega_a$ is an (unknown) differential $(\D-1)$-form on $\CM$.  Using Eq.~(\ref{eefive}) in coordinates, we can locally rewrite the geometrical Hamilton-Jacobi equation as
\be
w^\ast\theta=d(Z-w^\ast\sigma)\equiv dS.
\label{eenine}
\ee
Now, $w$ can be algebraically solved for in terms of $S$ and its derivatives, and the Hamilton-Jacobi equation (\ref{eeone}) follows.
\smallskip

For first order Lagrangians, it has been proved in Ref.~\cite{gs} that, given a $w$ that solves Eq.~(\ref{eeseven}), any field configuration embedded to $w$ (see Ref.~\cite{gs} for definition) solves the corresponding Lagrange equations. It is easy to show that the proof given in Ref.~\cite{gs} goes through in the gravitational case as well.  Now, the conditions for a metric field $\bar h^{ab}$ to be embedded to $w$ that solves the Hamilton-Jacobi equation (\ref{eeone}) are precisely Eqs.~(\ref{eethree}).  Hence, we have proved that any solution of Eqs. (\ref{eeone},\ref{eethree}) solves the Einstein equations.
\medskip

4. To establish the claimed equivalence between the Hamilton-Jacobi and Lagrange formulations, 
we shall now show that any solution of the Einstein equations can be embedded into a solution of the Hamilton-Jacobi equation.
\smallskip

In the case of regular first-order lagrangians, the existence of an appropriate solution of the corresponding Hamilton-Jacobi equation is proved in Ref.~\cite{gs} locally in space-time, without actually constructing it.  With some slight modifications, this proof might be extended to gravity.  Nevertherless, we shall take another way, and prove the equivalence theorem by constructing explicitly to any solution of the Einstein equations a solution of the Hamilton-Jacobi equation it is embedded to.
\smallskip

To this goal, let us consider an arbitrary $\bar h^{ab}(x)$ that solves the Einstein equations, and take the Ansatz
\be
S_0^a=h^{bc}Q_{bc}^a,
\label{eeten}
\ee
where $Q_{bc}^a$ are functions of $x^d$ only, constructed with the use of our metric $\bar h^{ab}$ as in Eq.~(\ref{eesix}), $\Gamma$ now representing the Christoffel symbols of $\bar h^{ab}$.  (Note that there is no bar over the $h^{bc}$ on the right-hand side of Eq.~(\ref{eeten}).) Plunging this into the Hamilton-Jacobi equation (\ref{eeone}) and the embedding conditions (\ref{eethree}), we obtain the following set of conditions:
\bea
\frac{\p Q_{bc}^a}{\p x^a}&=&-\left\{Q_{bd}^eQ_{ec}^d+\frac{1}{(1-\D)}Q_{bd}^dQ_{ce}^e\right\},
\label{eeeleven}\\
\frac{\p\bar h^{bc}}{\p x^a}&=&\bar h^{d(b}Q_{ad}^{c)}+\frac{1}{(1-\D)}\delta_a^{(b}\bar h^{c)d}Q_{de}^e.
\label{eetwelve}
\eea
It is easy to show that these conditions represent what we have called the Hamilton-De~Donder equations for gravity, with the Hamiltonian equal to
\be
\CH=h^{bc}\left\{Q_{bd}^aQ_{ac}^d+\frac{1}{(1-\D)}Q_{bd}^dQ_{ac}^a\right\}.
\nonumber
\ee
Because $\bar h^{ab}$ solves the Einstein equations, the couple $(\bar h^{ab},Q_{bc}^a)$ solves their Hamilton-De~Donder version (\ref{eeeleven}--\ref{eetwelve}) (regularity of $\CL$), thus proving that the Ansatz (\ref{eeten}) yields a solution of the Hamilton-Jacobi equation appropriate for obtaining the solution $\bar h^{ab}$ of the Einstein equations from the Hamilton-Jacobi theory via the procedure outlined above.
\smallskip

Note that, as a byproduct of this reduction of Eqs.~(\ref{eeone},\ref{eethree}) to Eqs.~(\ref{eeeleven},\ref{eetwelve}), the Hamilton-De~Donder description of gravity can be thought of as a part of the Hamilton-Jacobi theory, this interesting property being specific to Einstein gravity.
\medskip

5. To show that any solution of the Einstein equations can be obtained from the Hamilton-Jacobi theory even on topologically non-trivial manifolds, we shall again make use of the Ansatz (\ref{eeten}).  From the transformation properties of Christoffel symbols, we can infer transformation properties of $Q_{bc}^a$, and consequently transformation properties of $h^{bc}Q_{bc}^a$.  It is then easy to observe that the Ansatz is preserved with respect to coordinate transformations, thereby comprising a globally well-defined action form $Z$.
\smallskip

After plunging the Ansatz (\ref{eeten}) into Eq.~(\ref{eetwo}), we reach the surprising conclusion that the global action form $Z$ is identically equal to zero:
\be
Z^a=S_0^a-\frac{\p S_0^a}{\p h^{bc}}h^{bc}=h^{bc}Q_{bc}^a-h^{bc}Q_{bc}^a=0,
\label{eethirteen}
\ee
independently of the particular solution of the Einstein equations we have picked up!  It in particular means that we have obtained a universal solution of the Hamilton-Jacobi equation, into which any solution of the Einstein equations can be embedded.  This result is nevertheless of little practical importance for actually solving the Einstein equations, since the correspondence (\ref{eetwo}) is not one-to-one, and $w$ cannot be reconstructed from $Z$ itself.  Despite this fact, it seems interesting that from the viewpoint of the global Hamilton-Jacobi theory, the description of gravity significantly simplifies.  These facts bring to mind Witten's suggestions \cite{witten,lpw} on the topological phase of quantum gravity, with the fundamental Lagrangian for gravity being possibly zero, or a topological invariant.
\smallskip

Returning to the geometry of Section~(3), note that we have actually solved the problem of equivalence between the two versions of the Hamilton-Jacobi equation, which might have superficially looked like a complicated cohomological problem.  Whereas in the beginning of this section we showed that the strong form of the Hamilton-Jacobi equation, namely $[w^\ast\rho]=0$ as an element of the cohomology group $H^\D(\CM,\R)$, suffices to produce all solutions of the Einstein equations, we have now observed that an even stronger form of the Hamilton-Jacobi equation, namely $w^\ast\rho=0$, is actually sufficient.
\medskip

6. To conclude:  We have established a covariant formulation of classical gravity, which extends naturally the Hamilton-Jacobi theory of mechanics to pure gravity, and is equivalent to the standard Lagrange approach on any manifold on which the pure gravity may exist.
\smallskip

One may wonder whether an analogous Hamilton-Jacobi description exists when matter is coupled to gravity.  Actually, the covariant Hamilton-Jacobi framework presented here can be extended to the case of gravity coupled to electromagnetism; the resulting Hamilton-Jacobi equation has some interesting properties that amount to an amusing simplification of the theory at the level of its Hamilton-Jacobi description.  These results as well as some more details of the Hamilton-Jacobi theory for pure gravity presented here will appear elsewhere \cite{hjem}.
\smallskip

To be able to study possible implications of the framework to quantum gravity, it is first necessary to clarify connections between the covariant Hamilton-Jacobi equation presented above, and the conventional, ADM Hamiltonian formulation with its own Hamilton-Jacobi equation \cite{peres}.  Then, one might wonder whether the covariant Hamilton-Jacobi equation may offer some hints on the version of quantum gravity it is a WKB approximation of.  There is at least one area in which some concrete calculations along these lines could be made.  In the context of the considerable progress made recently in quantization of gravity in $2+1$ dimensions (see Ref.~\cite{carlip} and references therein), it might be illuminating to study connections between our formulation of Hamilton-Jacobi theory, perfectly valid in $2+1$ dimensions, and the full-fledged quantum theory, with the aim to learn whether the Hamilton-Jacobi equation may give some hints about quantum theory, and to identify the relations of the Hamilton-Jacobi framework to other semiclassical descriptions of quantum gravity.

\vfill\pagebreak
\bibliographystyle{JHEP}
\bibliography{hj}
%%%%%%%%%%%%%%%%%%%%%%%%%%%%%%%%%%%%%%%%%%%%%%%%%%%%%%%%%%%%%%%%%%%%%%%%%%%%%%%
\begin{center}
  \hglue-.5in
\includegraphics[angle=0,width=7in]{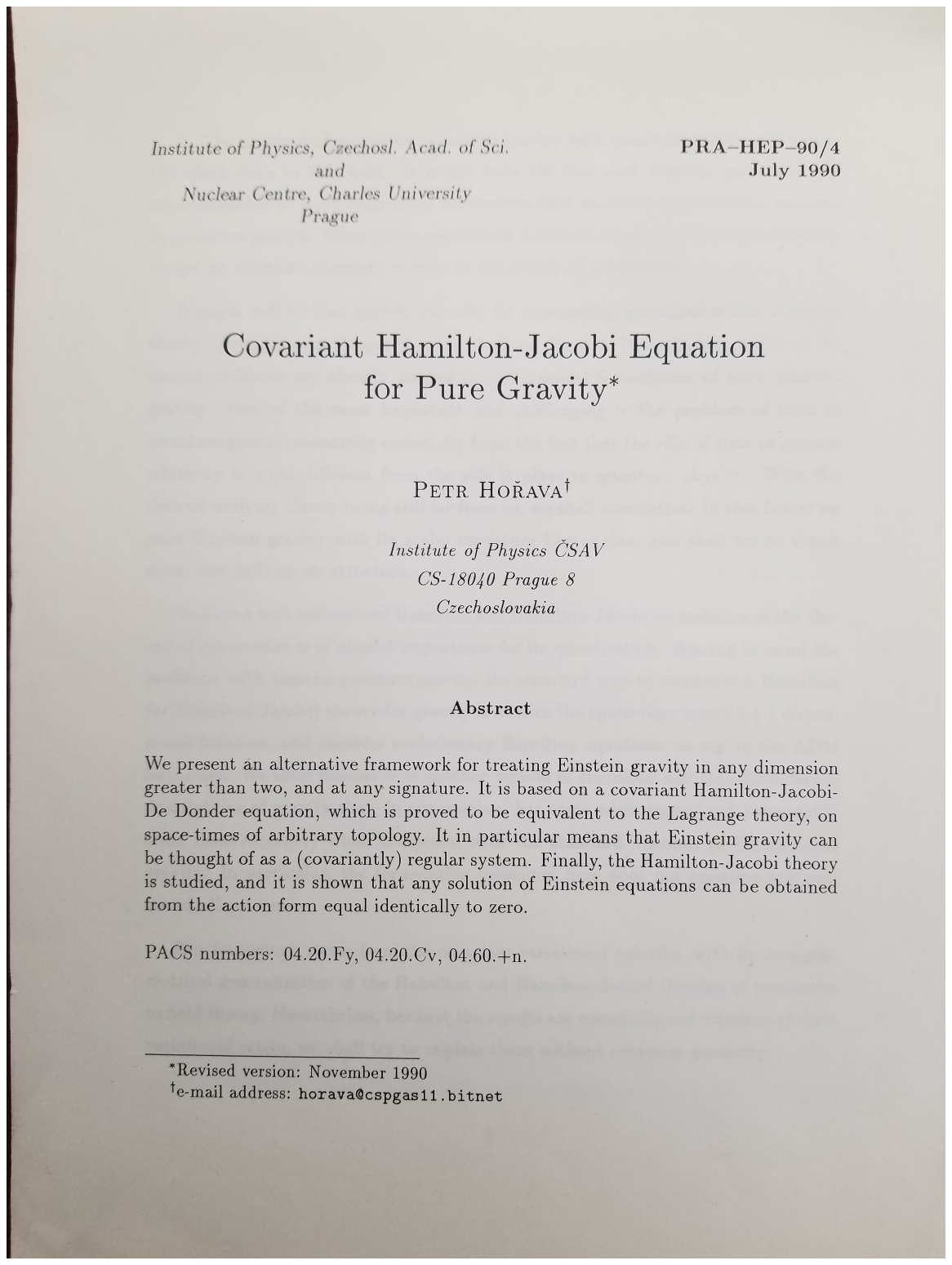}
        \vfill\break
  \hglue-.5in
\includegraphics[angle=0,width=7in]{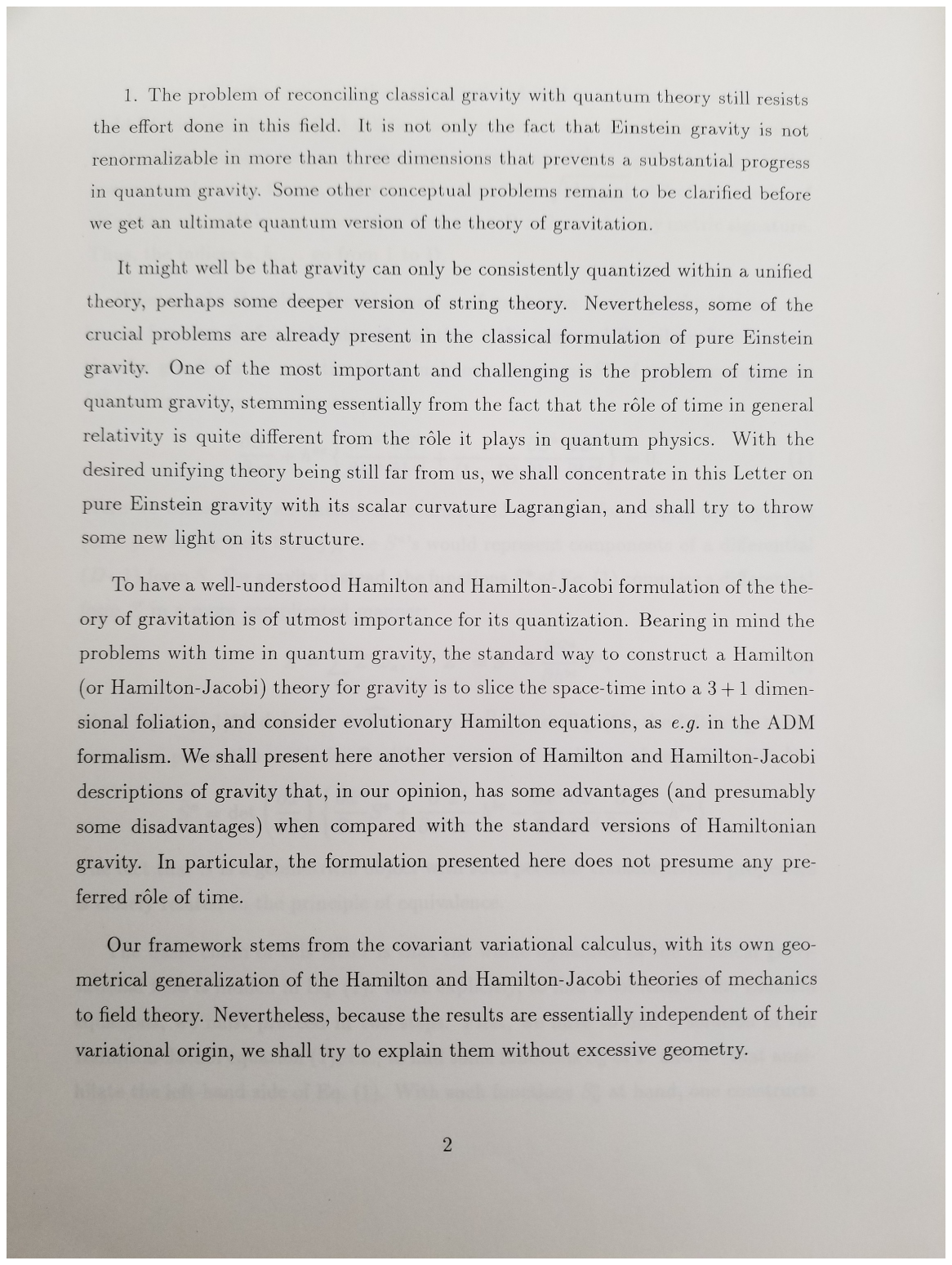}
        \vfill\break
  \hglue-.5in
\includegraphics[angle=0,width=7in]{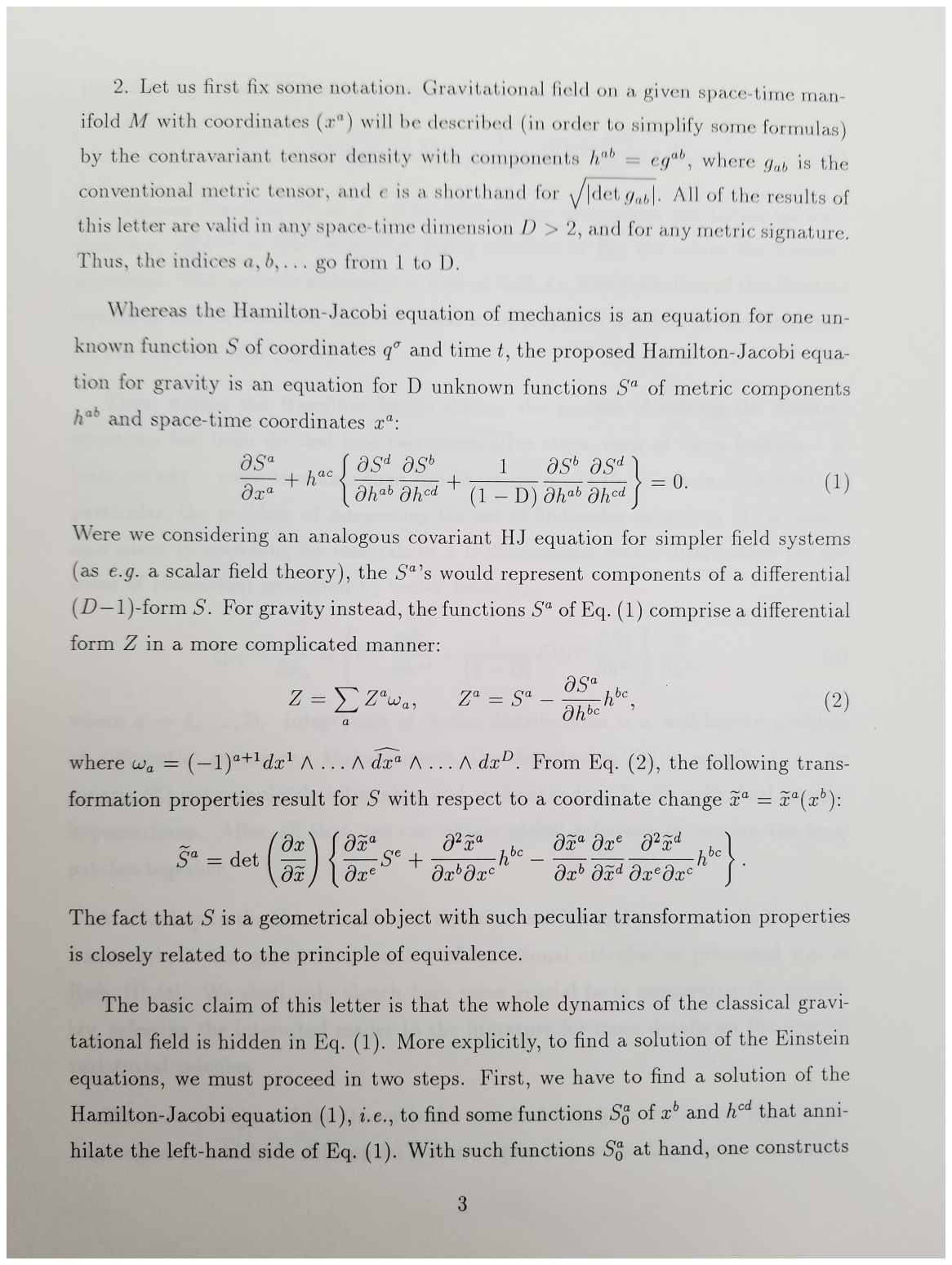}
        \vfill\break
  \hglue-.5in
\includegraphics[angle=0,width=7in]{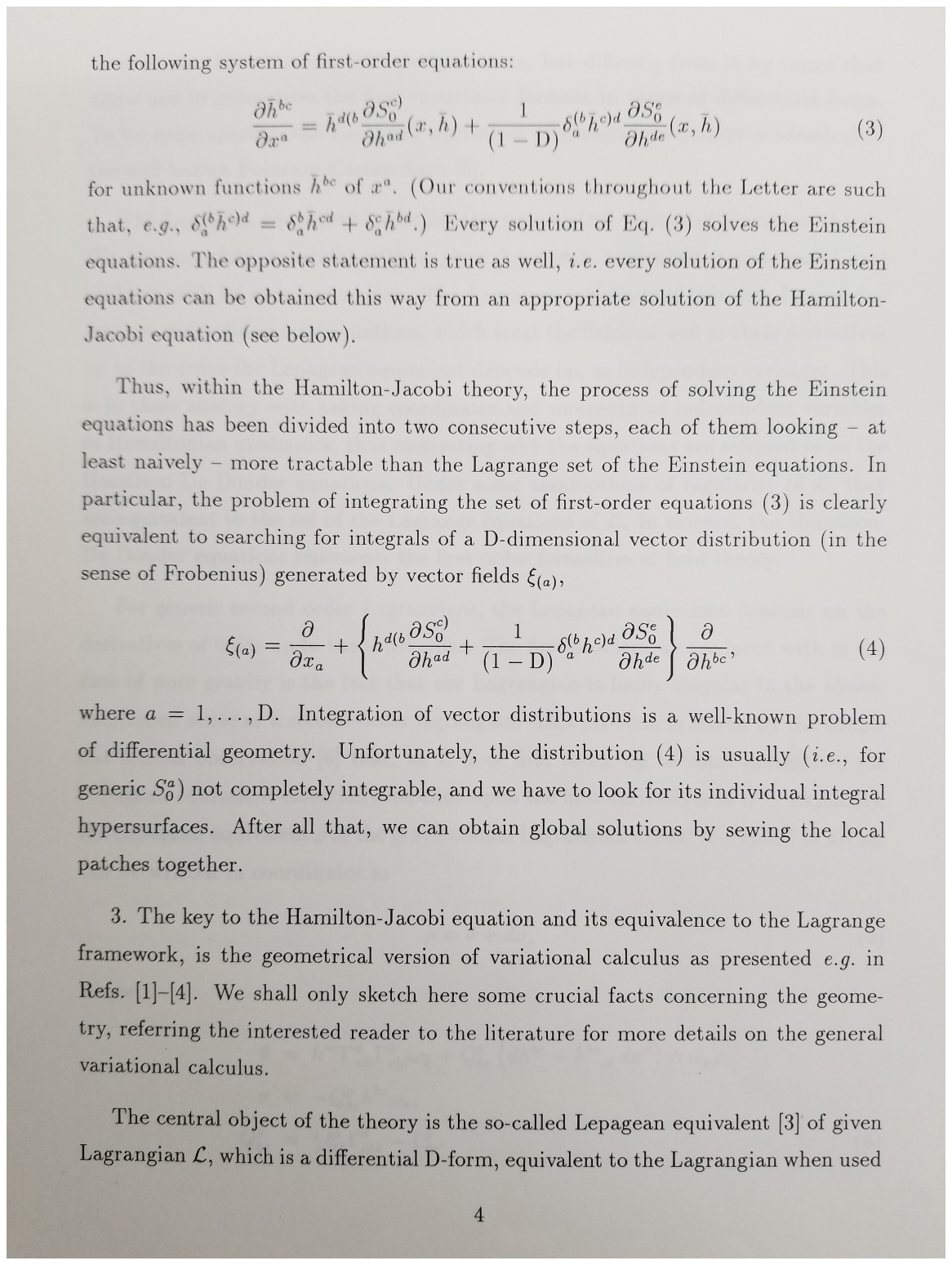}
        \vfill\break
  \hglue-.5in
\includegraphics[angle=0,width=7in]{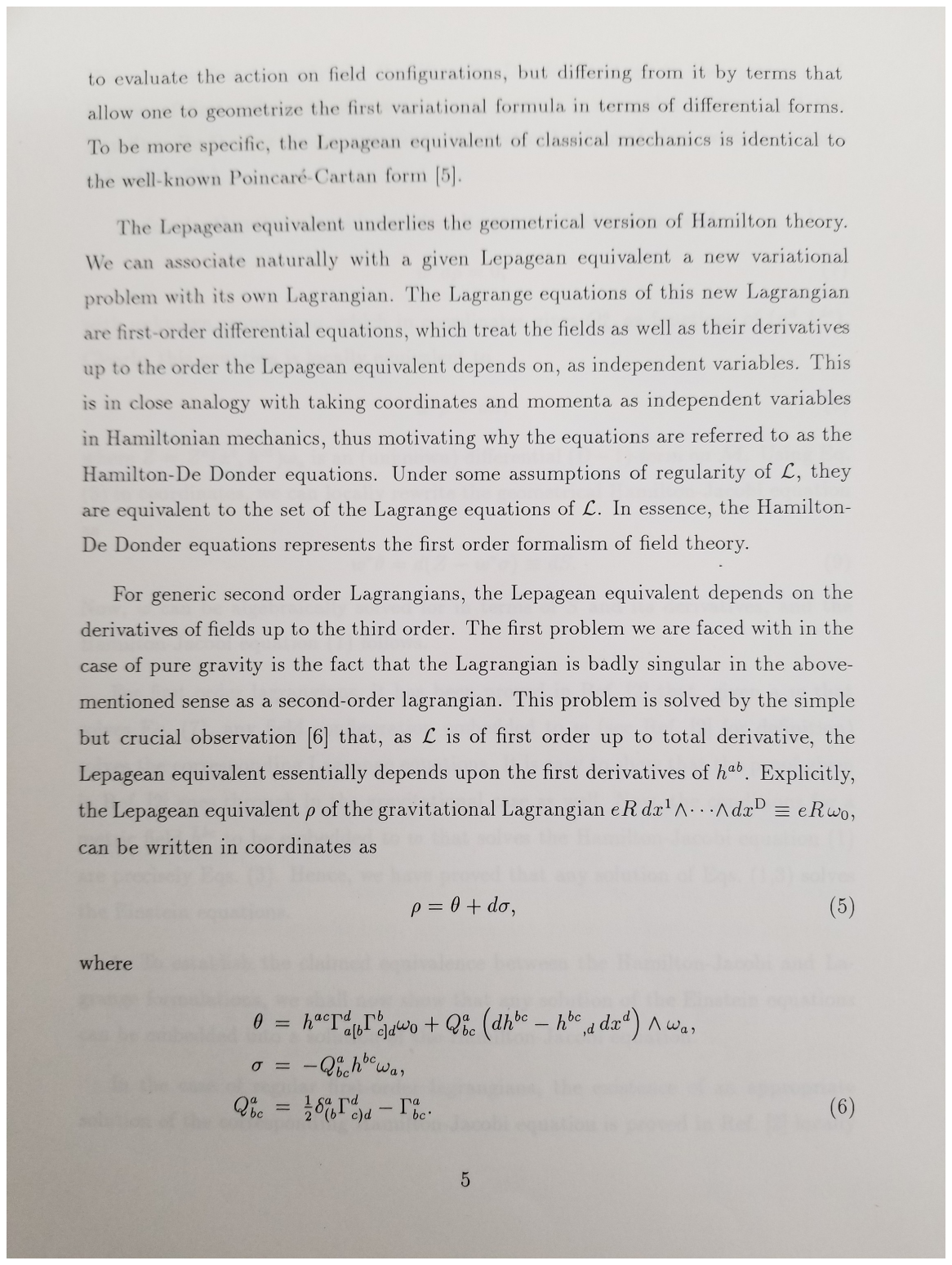}
        \vfill\break
  \hglue-.5in
\includegraphics[angle=0,width=7in]{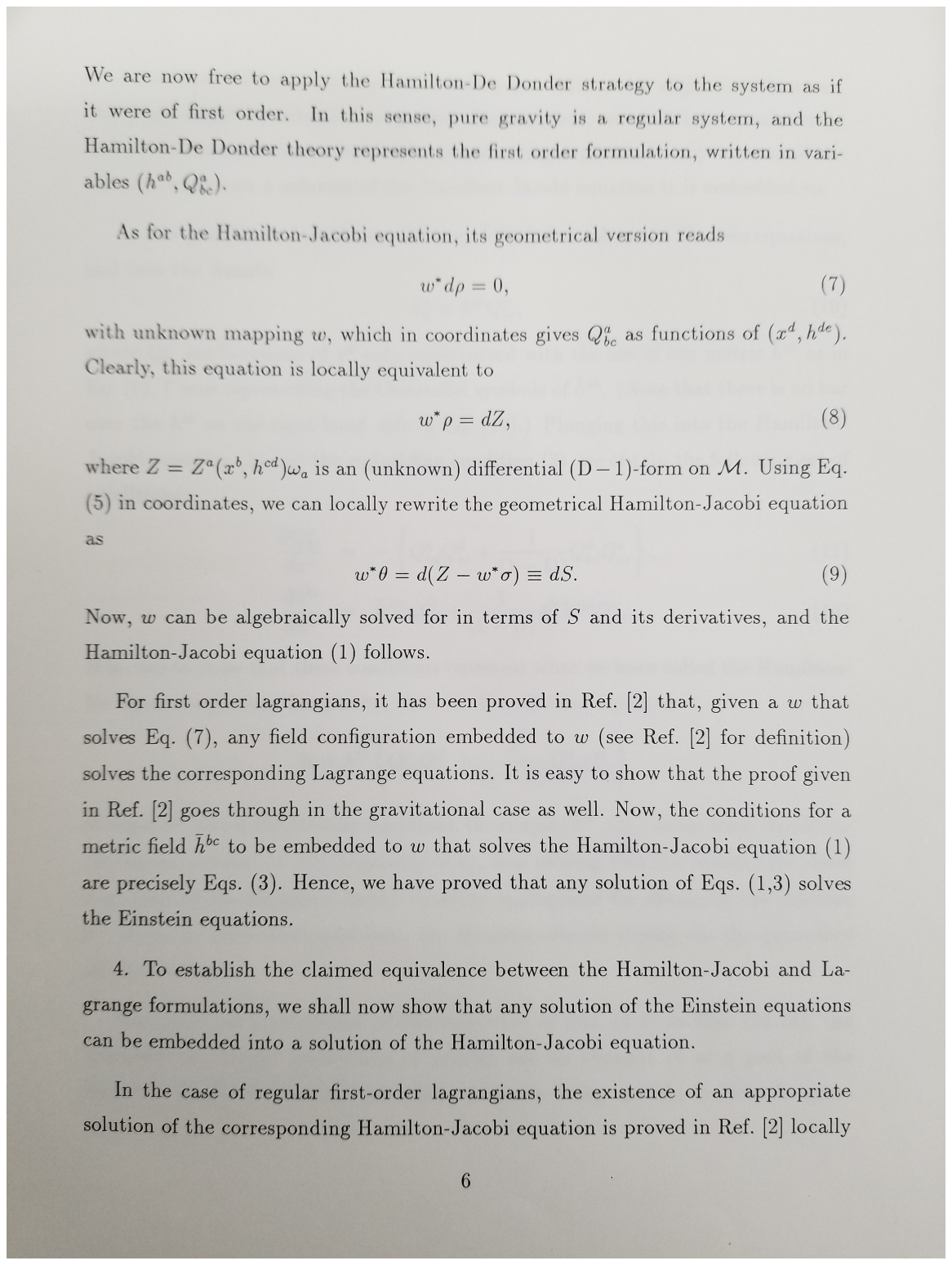}
        \vfill\break
  \hglue-.5in
\includegraphics[angle=0,width=7in]{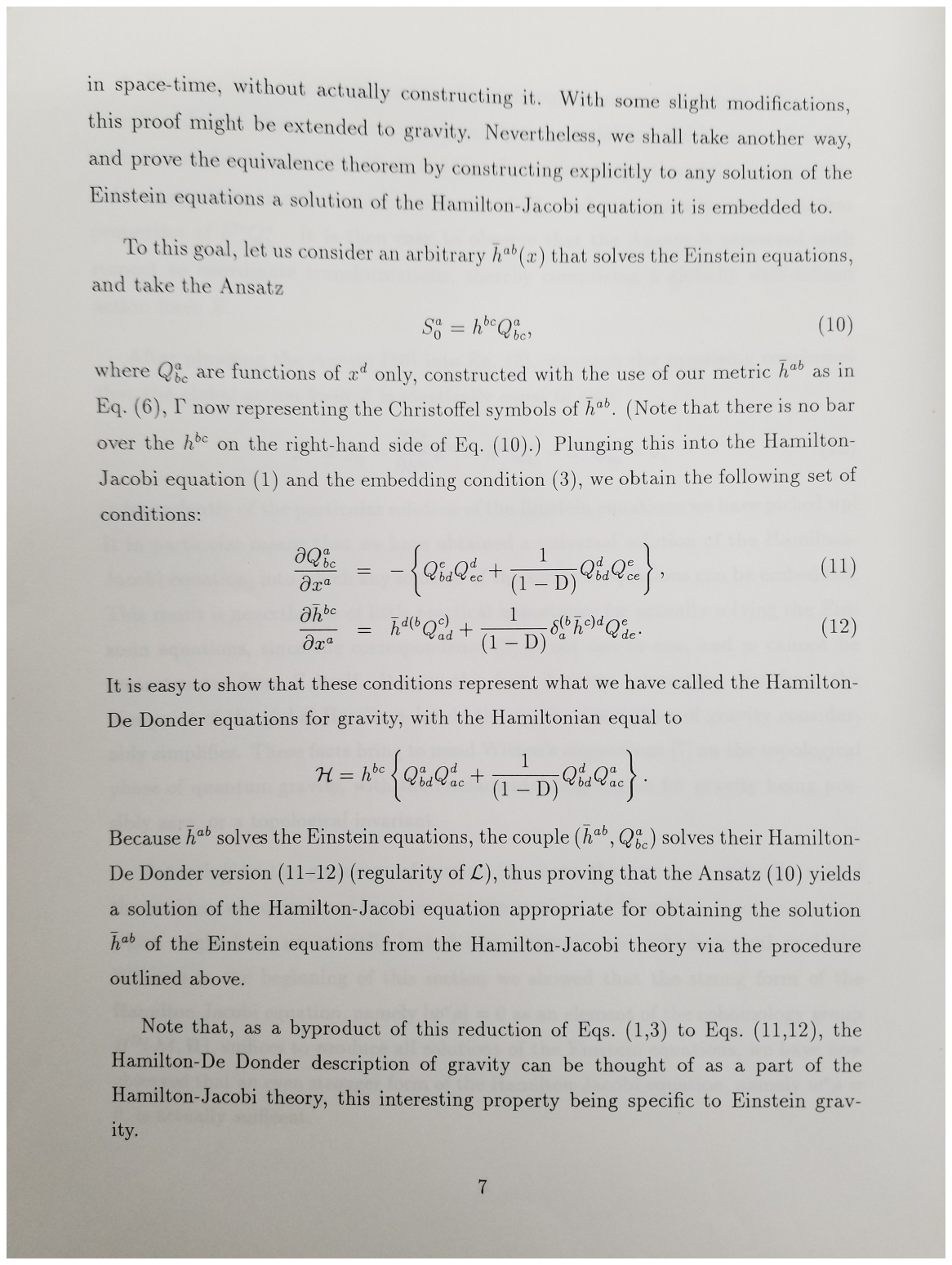}
        \vfill\break
  \hglue-.5in
\includegraphics[angle=0,width=7in]{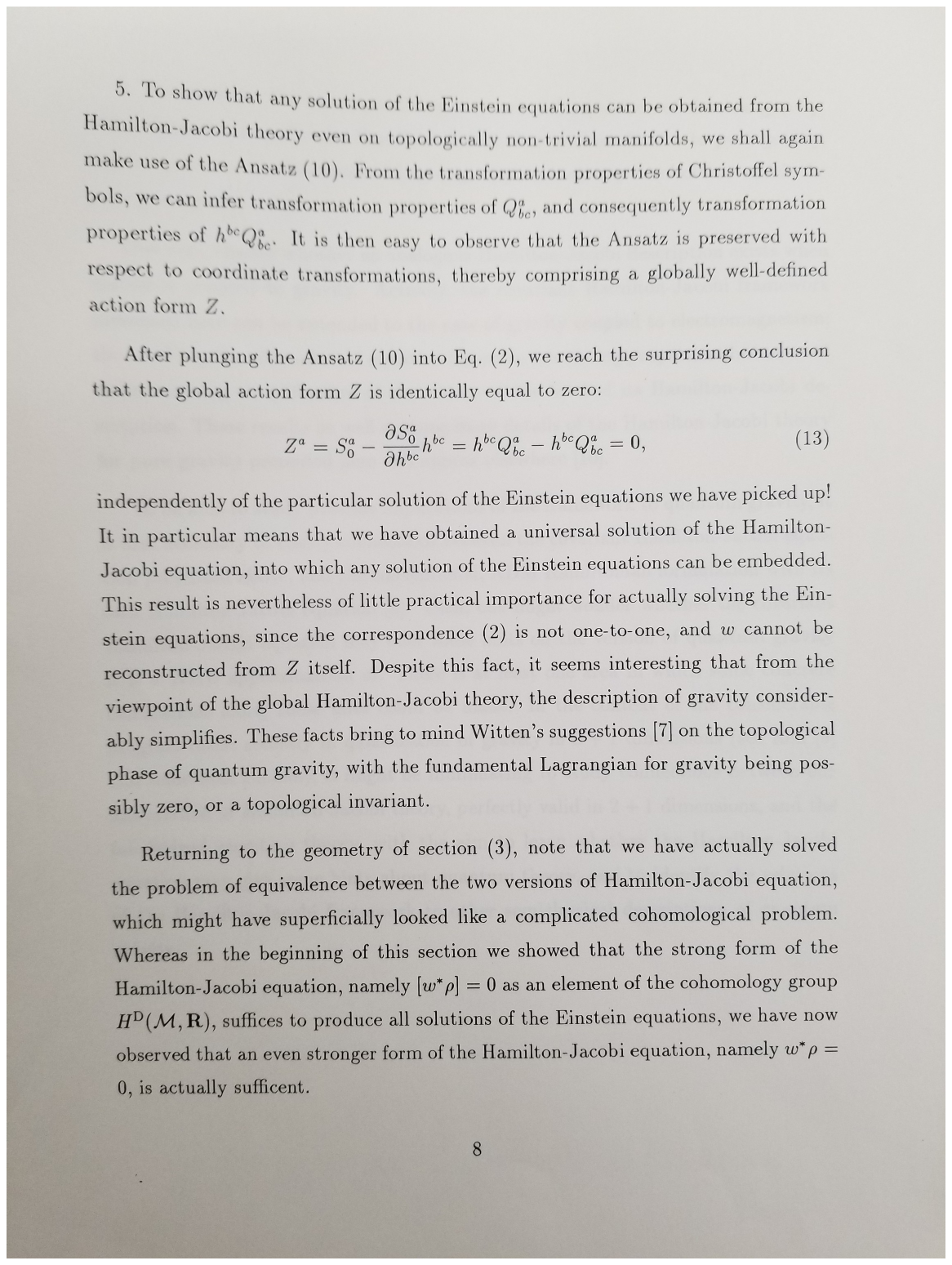}
        \vfill\break
  \hglue-.5in
\includegraphics[angle=0,width=7in]{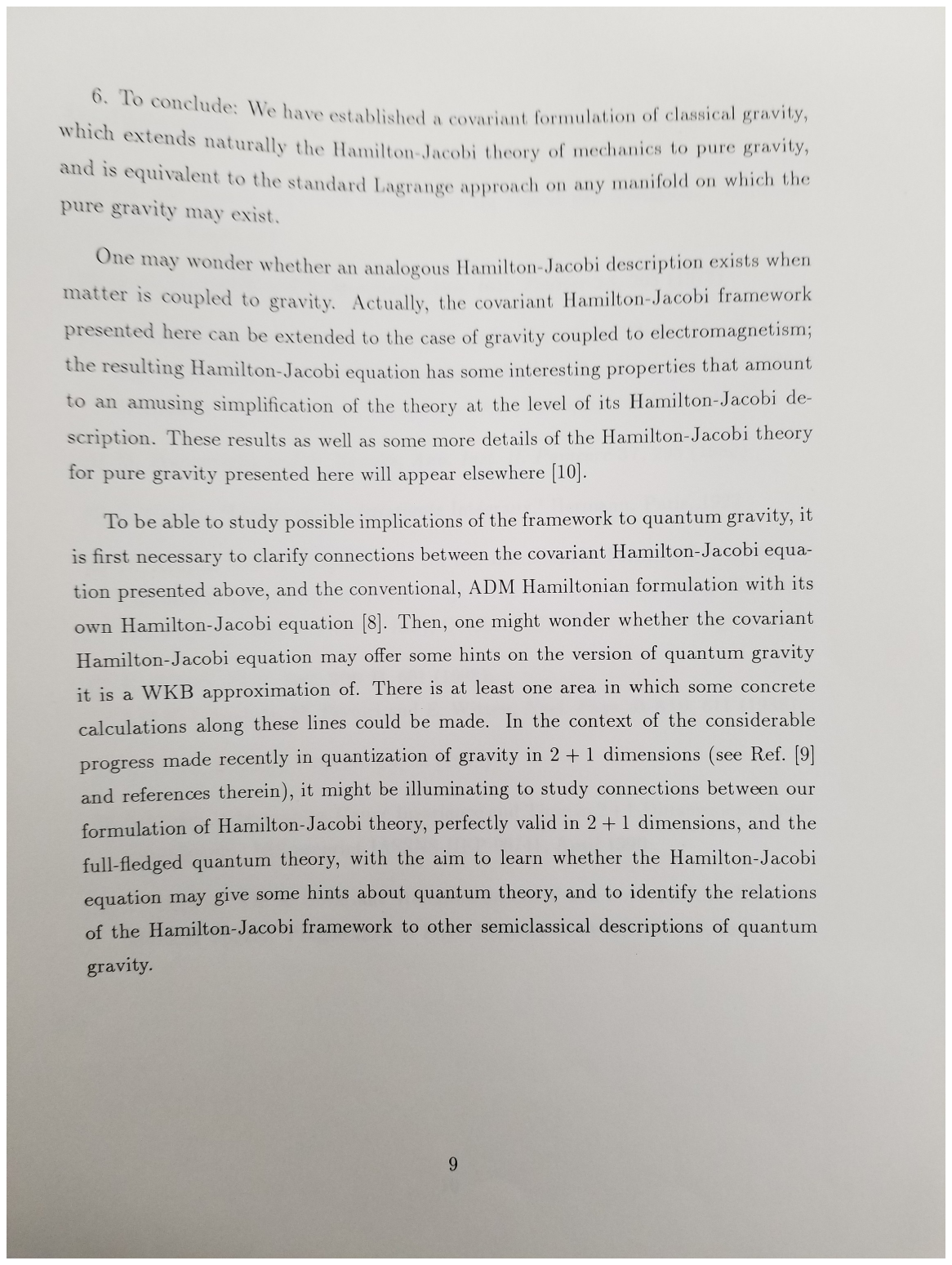}
        \vfill\break
  \hglue-.5in
\includegraphics[angle=0,width=7in]{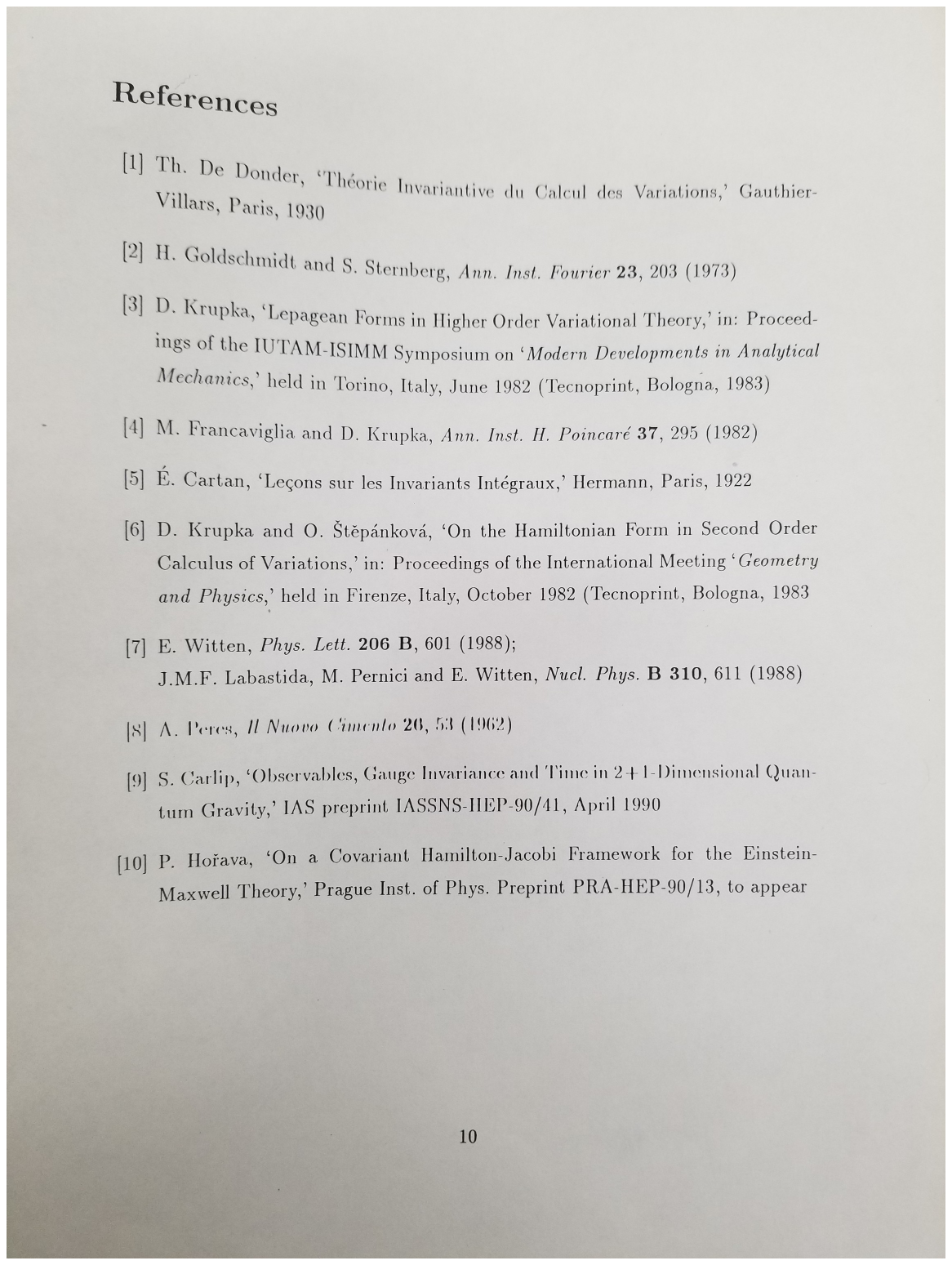}
        \vfill\break
\end{center}
\end{document}